\documentstyle[11pt,paspconf,epsf,psfig,twoside]{article}

\newcommand{\SS}{\scriptscriptstyle}
\newcommand{\Rl}{R_{\SS L1}}
\begin{document}


\title{Accretion Stream Mapping}


\author{Sonja Vrielmann$^1$, Axel D.\ Schwope$^2$}
\affil{$^1$Dept.\ of Astronomy, University of Cape Town, South Africa\\
$^2$Astrophysikalisches Institut Potsdam, Germany}

\begin{abstract}
We present a new mapping algorithm, the {\em Accretion Stream
Mapping}, which uses the complete emission-line light curve to derive spatially
resolved intensity distributions along the stream on a surface created
as a duodecadon shaped tube. We successfully test this method on
artificial data and then applied it to emission line light curves in
H$\beta$, H$\gamma$ and He~II~4686 of the magnetic CV HU~Aqr.  We find
Balmer emission near the threading point in the stream facing the
white dwarf and Helium emission all over the magnetic part of the
stream.
\end{abstract}


\section{Introduction}
The light curves of eclipsing magnetic CVs contain a wealth of information
about the objects and in particular about the stream. The emission
from various parts of the system (the white dwarf, the stream and the
secondary), however, are superposed onto each other making it usually
difficult to distinguish the various sources. The identification of
the location of emission in the stream, one the other hand, may help
in solving problems concerning the interaction of the plasma in the
stream with the magnetic field of the white dwarf.

Assuming one can model or neglect emission from other parts of the
system, the light curve is dominated by emission from different parts
along the stream. Depending on where exactly in the stream material is
heated and emitting radiation, structures will appear in the light
curve at certain phases and particularly (however not only) in the
shape of eclipse profile.

Analogically to the eclipse mapping of non-magnetic CVs (Horne 1985),
a tomographic method which allows one to reconstruct the intensity
distribution in the accretion disc by fitting the eclipse profile, we
apply a similar maximum entropy method (MEM) to magnetic CVs. This
technique shall distinguish the locations of emission, e.g.\ near the
threading point where the stream material couples to the magnetic
field of the white dwarf or near the white dwarf.

A similar method was applied by Hakala (1995) and later by Harrop-Allin
et\,al.\ (1997 and these proceedings) using a genetic algorithm
to derive the intensity distribution along the stream
using eclipse profiles measured in broad-band filters and more or less
simplified accretion stream geometries. In analysing U and UBVR light
curves, respectively, of HU~Aqr, they find that the intensity
increases towards the white dwarf in the stream with a second local
maximum near the inner Lagrangian point in the high state
(Harrop-Allin et\,al.).
 
\section{The method and application to artificial data}

\begin{figure}
\hspace*{1cm}
\begin{minipage}{5cm}
\psfig{file=vrielmann1.epsi,width=5cm}
\end{minipage}
\hspace{1cm}
\begin{minipage}{4cm}
\psfig{file=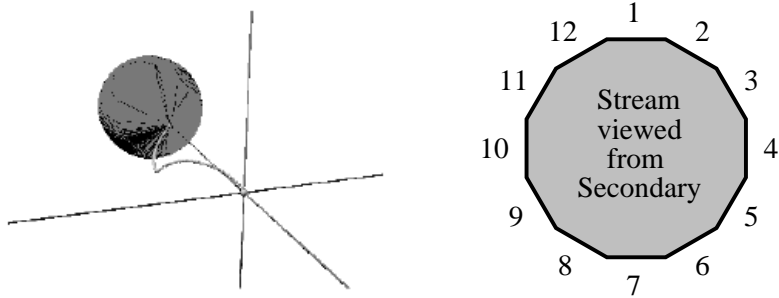,width=4cm}
\end{minipage}
\caption{{\em Left:} The geometry of the system (with coordinate axis
centred in the white dwarf). {\em Right:} The duodecadon shaped stream
crosssection as viewed from the inner Lagrangian point L$_1$ on the
secondary.
\label{geo}}
\end{figure}

The stream we use for our {\em Accretion Stream Mapping} techniques
follows a ballistic
trajectory down to the threading region and then couples
to a dipolar field line (Fig.~\ref{geo}).
It is assumed to be a non-transparent 12 sided tube\footnote{Pixel 1
is always facing upwards, i.e.\ the area normal is parallel or closest
to the orbital axis. Pixel 4 facing {\em outwards} and pixel 10 {\em
inwards}.}.  While the system is rotating, the pixel appear under
varying angles and are eclipsed at certain phases either by the stream
itself or by the secondary. This justifies the use of the {\em full
observed light curve}.


For the reconstruction of real accretion streams, four parameters
(mass ratio, inclination angle, location of the threading point,
orientation of the magnetic axis) have to be determined which describe
the geometry of the stream. 
The parameters, however, often can be determined quite well from
{\em Doppler Tomography} (see Schwope, these proceedings).

We constructed a stream geometry for a CV with the {\em system parameters}:
$i=85$, $q=0.2$, a rather strong magnetic field and an axis almost
parallel to the orbital axis (20$^\circ$ away and 45$^\circ$ from the
binary axis). Between the radii 0.4 and 0.5, shortly before the
connecting point, we placed a bright spot facing the white dwarf
otherwise the stream emits only 10\% of the spot intensity. For
this artificial stream we calculated the light curve
(Fig.~\ref{test_lcv}, {\em left}) and analysed it with the {\em Accretion
Stream Mapping} method.

\begin{figure}
\hspace*{1cm}
\begin{minipage}{5cm}
\psfig{file=vrielmann3.epsi,width=5cm}
\end{minipage}
\hspace*{1cm}
\begin{minipage}{5cm}
\psfig{file=vrielmann4.epsi,width=5cm}
\end{minipage}
\caption{{\em Left:} The light curve of the artificial intensity
distribution with the fit. {\em Right:} The original (small dots) and
reconstructed (larger dots) radial intensity profile on the stream
surface. The abzissa gives the distance from the white dwarf and the
ordinate the intensity distribution offset by 0.2 for each pixel
row. The short row of dots at 1.4 intensity units belongs to the
original intensity distribution in row 10. The vertical dotted line at
$0.43 \Rl$ denotes the location where the magnetic part of the stream
starts.\label{test_lcv}}

\vspace{2ex}
\hspace*{1.5cm}
\begin{minipage}{4.5cm}
\psfig{file=vrielmann5.epsi,width=4.5cm}
\end{minipage}
\hspace*{1cm}
\begin{minipage}{4.5cm}
\psfig{file=vrielmann6.epsi,width=4.5cm}
\end{minipage}
\caption{{\em Left:} The original intensity distribution: In the spot
the pixel have an intensity value of 1 (arbitrary units), while the
other pixel have an intensity value of 0.1. The horizontal dotted line
at pixel 65 denotes the location where the magnetic part of the stream
starts, the white dwarf is at the top, the secondary at the bottom.
{\em Right:} The reconstructed intensity distribution.
\label{test_grey}}
\end{figure}

Fig.~\ref{test_grey} shows gray-scale images of the original and
reconstructed intensity distributions. The {\em location of the spot} is
reconstructed very well, it is only smeared out due to the MEM
algorithm, leading as well to a different spot profile and lower spot
intensity (see also Fig.~\ref{test_lcv}, {\em right}). 

We tested this method also against picking wrong values for one to
three of the four system parameters values. The spot is
in some cases reconstructed, but (additional) spots at other locations
emerged occasionally.

These tests, however, show the advantage of using the full light
curve, since it indeed constrains the system parameters fairly
well. In comparison, Hakala (1995) found not much difference for his
reconstruction of the accretion stream for different sets of
system parameters.

\begin{figure}[t]
\hspace*{2.5cm}
\psfig{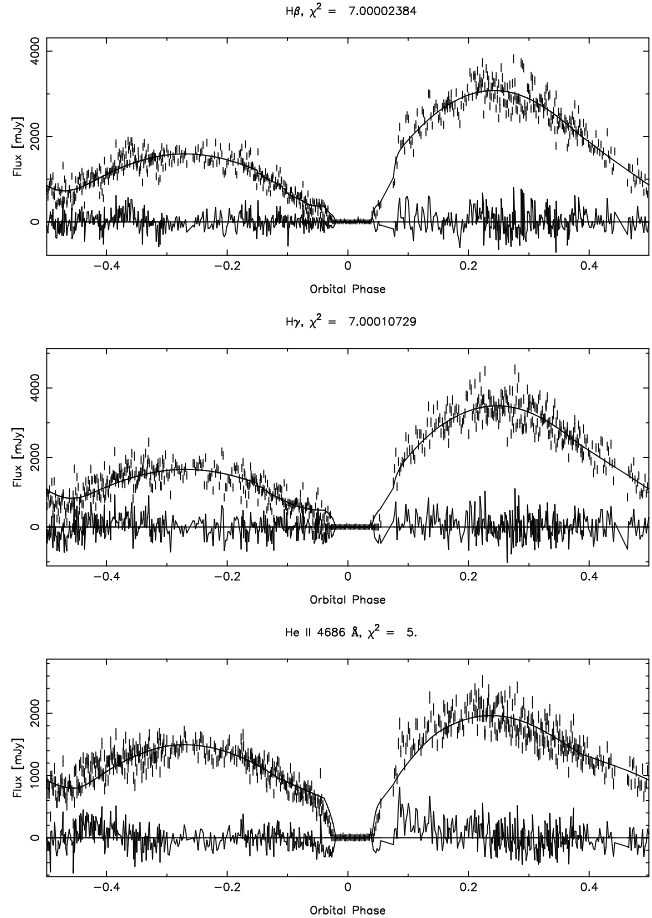}
\caption{The light curves in the emission lines H$\beta$, H$\gamma$ and
He~II~4686 with the fits. In the bottom of each plot the residuals are
given.\label{lcvs}}
\end{figure}
\nopagebreak

\section{Application to emission line data of HU~Aqr}

\begin{figure}[t]
\hspace*{2.5cm}
\psfig{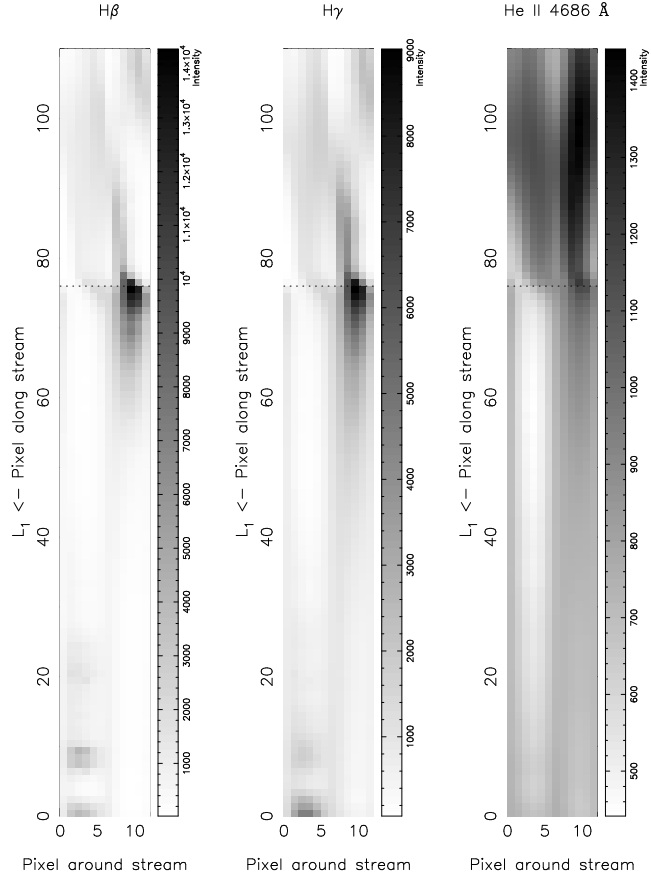}
\caption{The recontructed intensity distributions on the surface of
the accretion stream as gray-sclae plots.The horizontal dotted line
at pixel 76 denotes the threading location.}
\end{figure}
\nopagebreak

Since the analyses of broad band photometry would force us to include
absorption and reprocession effects, we applied it first only to light
curves in various emission lines and therefore also excluded the
white dwarf as a source of light. We used high time and high spectral
resolution spectra of the magnetic CV HU~Aqr in the high state taken at
Calar Alto, Spain in 1993 (Schwope, Mantel, Horne 1997).
The integrated line profiles of H$\beta$, H$\gamma$, and He~II4686 
were corrected for emission orginating
on the irradiated surface of the secondary. The remaining
emission is regarded to originate solely from the accretion stream;
the resulting light curves are shown in Fig.~\ref{lcvs}.

{\em Accretion Stream Mapping} was applied to these light curves using
the following system parameters: mass ratio $q = 0.25$, inclination $i
= 85^\circ$, field orientation as in the test (20$^\circ$, 45$^\circ$)
and location of the threading point as determined via {\em Doppler
Tomography} (Schwope, Mantel, Horne 1997).  The {\em Accretion
Stream Mapping} method reached very good fits (Fig.~\ref{lcvs}). The
somewhat high $\chi^2$'s of 5 and 7 are mainly due to the large
scatter of the individual points in the light curve partly because of
difficulties in identifying the NEL. The overall shape of the light
curve was reconstructed very
well, except for the eclipse profile. Unfortunately, during the egress
(which is determined by the orientation of the magnetic field) only a
few data points were retrieved.

The reconstructions show a clear bright spot for the Balmer lines at
the threading region facing the white dwarf. A similar bright spot was
also found by Hakala (1995), however, his geometry did not allow to
distinguish between the illuminated and the non-illuminated sides of
the stream. At this point, the matter is drawn out of the orbital
plane, dissipation of kinetic energy is likely to occur resulting in
enhanced line emission. The location on the inner side of the stream
shows that the material must also be heated by the white dwarf and
dissipating this energy.

The intensity map in the He~II line appears much different from the
Balmer emission. Though, the light curve shows a worse S/N ratio,
clear differences in the light curve are seen which are reflected in
the intensity distribution: The difference between the maxima is less
pronounced for the Helium line and the eclipse depth and profile
differs significantly.

The reconstruction shows a much smoother intensity distribution than
the ones for the Balmer line. The side facing the white dwarf (pixel
rows 9 to 11) is everywhere bright. Also the magnetic part of the
stream appears bright. This is unexpected, because the white dwarf is
not able to directly irradiate (all) those regions. It appears that in
the magnetic part of the stream the line opacity is not negligible
(otherwise the stream would show no variation in intensity) but
significantly lower than in the ballistic part. However, since the
fits around eclipse are not perfect, yet, this conclusion is only
preliminary.

A similarly bright magnetic stream was found by Hakala and
Harrop-Allin et\,al.\ using broad band photometry.  However, {\em
Doppler Tomography} suggest a much more prominant ballistic stream,
especially in the high state (Schwope et\,al.). Clearly, one needs to
take into account both methods, {\em Doppler Tomography} and {\em
Accretion Stream Mapping}, since Doppler maps yield only the velocity
of the material, not the spatial location, and emission structures may
be smeared out significantly in velocity space. We intend, therefore,
to build a consistent picture including both techniques.


\acknowledgments SV thanks the south african FRD for funding a postdoc
fellowship.

%
%

\end{document}